\documentclass[amsmath,amssymb,twocolumn,prd,floatfix,showpacs, nofootinbib]{revtex4-1}
\usepackage{tabularx}  
\usepackage{bm, color} 
\usepackage{overpic,subfigure} 
\usepackage{multirow}
\usepackage{array}
\usepackage{dcolumn} 
\usepackage[symbol]{footmisc}
\usepackage{epstopdf}
\usepackage[normalem]{ulem}
\usepackage[T1]{fontenc}   
\usepackage{lmodern}        

\RequirePackage{xspace}
\newcommand{\gev}{\ensuremath{\mathrm{\,Ge\kern -0.1em V}}\xspace}
\newcommand{\mev}{\ensuremath{\mathrm{\,Me\kern -0.1em V}}\xspace}
\newcommand{\mevcc}{\ensuremath{{\mathrm{\,Me\kern -0.1em V\!/}c^2}}\xspace}

\renewcommand{\eqref}[1]{\textcolor{blue}{(\ref{#1})}}

\def\piz        {\ensuremath{\pi^0}\xspace}

\def\fz#1       {\ensuremath{f_0({#1})}\xspace}

\def\besiii     {BESIII\xspace}

\usepackage{hyperref}
\hypersetup{colorlinks = true,
            linkcolor = blue,
            urlcolor = blue,
            citecolor = blue,
            breaklinks=true,
            pdfstartview=Fit}
\begin{document}


\newcommand{\BESIIIorcid}[1]{\href{https://orcid.org/#1}{\hspace*{0.1em}\raisebox{-0.45ex}{\includegraphics[width=1em]{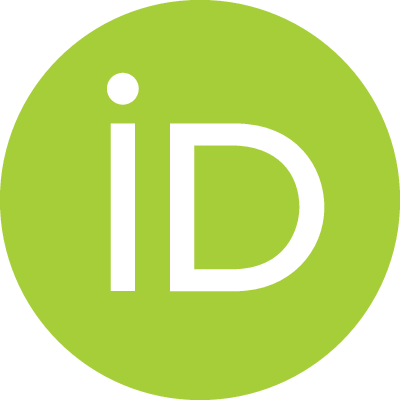}}}}

\title{\boldmath Observation of $\psi(3686)\to \gamma \eta(1405)$ via $\eta(1405)\to f_0(980)\pi^0$}

\author{
\begin{center}
M.~Ablikim$^{1}$, M.~N.~Achasov$^{4,d}$, P.~Adlarson$^{77}$, X.~C.~Ai$^{82}$, R.~Aliberti$^{36}$, A.~Amoroso$^{76A,76C}$, Q.~An$^{73,59,b}$, Y.~Bai$^{58}$, O.~Bakina$^{37}$, Y.~Ban$^{47,i}$
% ...
\\
\vspace{0.2cm}
(BESIII Collaboration)\\
\vspace{0.2cm} {\it
$^{1}$ Institute of High Energy Physics, Beijing 100049, People's Republic of China\\
% ...
$^{4}$ Budker Institute of Nuclear Physics SB RAS (BINP), Novosibirsk 630090, Russia\\
% ...
$^{36}$ Johannes Gutenberg University of Mainz, Johann-Joachim-Becher-Weg 45, D-55099 Mainz, Germany\\
$^{37}$ Joint Institute for Nuclear Research, 141980 Dubna, Moscow region, Russia\\
% ...
$^{47}$ Peking University, Beijing 100871, People's Republic of China\\
% ...
$^{58}$ Southeast University, Nanjing 211100, People's Republic of China\\
$^{59}$ State Key Laboratory of Particle Detection and Electronics, Beijing 100049, Hefei 230026, People's Republic of China\\
% ...
$^{73}$ University of Science and Technology of China, Hefei 230026, People's Republic of China\\
% ...
$^{76}$ University of Turin and INFN, (A)University of Turin, I-10125, Turin, Italy; (B)University of Eastern Piedmont, I-15121, Alessandria, Italy; (C)INFN, I-10125, Turin, Italy\\
$^{77}$ Uppsala University, Box 516, SE-75120 Uppsala, Sweden\\
% ...
$^{82}$ Zhengzhou University, Zhengzhou 450001, People's Republic of China\\

\vspace{0.2cm}
$^{b}$ Deceased\\
% ...
$^{d}$ Also at the Novosibirsk State University, Novosibirsk, 630090, Russia\\
% ...
$^{i}$ Also at State Key Laboratory of Nuclear Physics and Technology, Peking University, Beijing 100871, People's Republic of China\\
% ...

}

\end{center}
}

\begin{abstract}

The decay $\psi(3686)\to\gamma\pi^+\pi^-\pi^0$ is studied using a sample of $(2712.4\pm14.3)\times10^6$ $\psi(3686)$ events collected with the BESIII detector.
The decay $\eta(1405)\to\pi^+\pi^-\pi^0$ is observed for the first time in $\psi(3686)$ decays via the intermediate state $f_0(980)$ and the product branching fraction $\mathcal{B}(\psi(3686)\to\gamma\eta(1405))\times\mathcal{B}(\eta(1405)\to f_0(980)\pi^0)\times \mathcal{B}(f_0(980)\to\pi^+\pi^-)$ is determined to be
 $(3.77\pm0.43\pm0.30)\times10^{-7}$,
where the first uncertainty is statistical and the second is systematic. 
The branching fraction of isospin-violating decay $\psi(3686)\to\gamma f_1(1285)\to\gamma f_0(980)\pi^0\to\gamma\pi^+\pi^-\pi^0$ is determined to be $ (7.36\pm2.25\pm2.36)\times 10^{-8}$ with statistical significance of $2.9\sigma$. 
Since no $\eta_c$ signal is evident in either the $\pi^+\pi^-\pi^0$ or $f_0(980)\pi^0$ mass spectrum, 
upper limits are set to be 
$\mathcal{B}(\psi(3686)\to\gamma\eta_c)\times\mathcal{B}(\eta_c\to\pi^+\pi^-\pi^0)<3.09\times10^{-7}$ and $\mathcal{B}(\psi(3686)\to\gamma\eta_c)\times\mathcal{B}(\eta_c\to f_0(980)\pi^0)\times\mathcal{B}(f_0(980)\to\pi^+\pi^-)<7.97\times10^{-8}$ at 90\% confidence level, respectively. 

\end{abstract}

\maketitle

\section{Introduction}

A state near 1440~MeV/$c^{2}$, first discovered in $p\overline{p}$ annihilation at rest
decaying to $\eta\pi^+\pi^-$~\cite{AMSLER}, was subsequently shown by experiments~\cite{MARK-III,DM2} to consist of two different pseudoscalar states: the $\eta(1405)$ and the $\eta(1475)$. The former  couples strongly to $a_0(980)\pi$ and $K\overline{K}\pi$, while the latter mainly decays to $K^*\overline{K}$.
Extensive theoretical and experimental efforts have been devoted to understanding the nature of the $\eta(1405)$ and $\eta(1475)$~\cite{Cheng:2023lov,Nakamura:2022rdd,Du:2019idk,Achasov:2018swa,Wu}.

The isospin violating process $\eta(1405)\to f_0(980)\pi^0$ was first observed in $J/\psi$ radiative decays~\cite{PhysRevLett.108.182001}, 
with the ratio  $\frac{\mathcal{B}(\eta(1405)\to f_0(980)\pi^0)}{\mathcal{B}(\eta(1405)\to a_0(980)\pi^0)}=(17.9\pm4.2)\%$, which is one order of magnitude
larger than the most recent measurement of the $a_0(980)-f_0(980)$ mixing intensity (0.4\%)~\cite{ref::mixing}.
To account for this unexpectedly large isospin violation and the anomalously narrow width of the $f_0(980)$, Wu $et~al.$ proposed that a triangle
singularity mechanism could play a crucial role in this process~\cite{wujiajun}.
Further theoretical and experimental studies are needed for a better understanding of the underlying dynamics.

The axial vector meson $f_1(1285)$ could be interpreted as a $K^*\bar{K}$ molecule~\cite{f1285_1} or a tetraquark state~\cite{f1285_2} , however, its nature is still need to be understood. 
The isospin symmetry breaking decay $f_1(1285)\to 3\pi$ can proceed through several different mechanisms, with the direct decay $f_1(1285)\to3\pi$ expected to be small. One main mechanism is the $f_1(1285)-a_1(1260)$ mixing by assuming that the universality can be extended to the axial meson sector~\cite{f1285_3}.
Another is the $f_1(1285)\to f_0(980)\pi$ decays through $f_1(1285)\to a_0(980)\pi$ via $a_0(980)-f_0(980)$ mixing. 
VES Collaboration~\cite{f1285_4} first observed the $f_1(1285)\to \pi^+\pi^-\pi^0$ and conclude that the decay caused presumably by the ``$a_0(980)-f_0(980)$'' mixing, but extra decay mechanisms are necessary.
In BESIII previous analysis of $J/\psi\to\phi\pi^0 f_0(980)$ with 1.3 billion $J/\psi$ events~\cite{f1285_5}, the evidence of $f_1(1285)$ is found in the $f_0(980)\pi^0$ mass spectrum.

A previous search for the decay $\eta(1405)\to f_0(980)\pi^0\to\pi^+\pi^-\pi^0$ was performed by BESIII using a data sample of $(447.9\pm2.9)\times10^6$ $\psi(3686)$ events
and the upper limit (UL) at the 90\% confidence level (CL) was set to be $\mathcal{B}^{\rm UL}(\psi(3686)\to\gamma\eta(1405)\to\gamma f_0(980)\pi^0\to\gamma\pi^+\pi^-\pi^0)=5.0\times10^{-7}$ \cite{ref::gushanwork}.
Currently, BESIII has accumulated $(2712.4\pm14.3)\times10^6$ $\psi(3686)$ events~\cite{ref::psip_event}, which is approximately five times larger than the previous dataset, and allows further investigation of the decay $\eta(1405)/f_1(1285)\to f_0(980)\pi^0$.

In addition, according to  perturbative quantum chromodynamics (QCD), the ratio of the branching fractions of $\psi(3686)$ and $J/\psi$ decaying to the same hadronic final state is predicted to be
\begin{equation}
    Q=\frac{\mathcal{B}_{\psi(3686)\to h}}{\mathcal{B}_{J/\psi\to h}}=\frac{\mathcal{B}_{\psi(3686)\to l^+l^-}}{\mathcal{B}_{J/\psi\to l^+l^-}}\approx (12.4\pm0.4)\%,
\end{equation}
commonly referred to as the ``12\% rule'' \cite{ref::12rule1,ref::12rule2,ref::12rule3}, which works reasonably well for many decay modes. However, significant violations of this rule have been observed by subsequent experiments~\cite{violation1, violation2}, particularly in the $\rho\pi$ final state. Recent reviews~\cite{rhopi1,rhopi2,rhopi3} indicate that current theoretical explanations remain unsatisfactory.  Whether or not the 12\% rule also holds for radiative decays~\cite{ref::gushanwork} is another motivation to study the decay $\psi(3686)\to\gamma \pi^+\pi^-\pi^0$.

The same data sample is also used to search for the isospin-violating process $\eta_c\to\pi^+\pi^-\pi^0$, which provides an important test of isospin symmetry.
The $\eta_c$, the lowest-lying $c\overline{c}$ pseudoscalar state, has attracted considerable theoretical and experimental attention since its discovery \cite{ref::etac_discover}. It decays primarily via $c\overline{c}$ annihilation into two gluons and is expected to have numerous hadronic decay modes into
two- or three-body hadronic final states, and many of them have been measured~\cite{pdg}.
Using a subsample of $\psi(3686)$ events, BESIII previously set an UL of $\mathcal{B}(\eta_c\to\pi^+\pi^-\pi^0) < 4.0\times10^{-4}$ at the 90\% CL~\cite{ref::gushanwork}, which can be improved significantly with the higher-statistics sample used in this analysis.

\section{BESIII DETECTOR AND MONTE CARLO SIMULATION}

The BESIII detector~\cite{Ablikim:2009a} records symmetric $e^+e^-$ collisions 
provided by the BEPCII storage ring~\cite{Yu:IPAC2016-TUYA01}
in the center-of-mass energy range from 1.84 to 4.95~GeV,
with a peak luminosity of $1.1 \times 10^{33}\;\text{cm}^{-2}\text{s}^{-1}$ 
achieved at $\sqrt{s} = 3.773\;\text{GeV}$. 
BESIII has collected large data samples in this energy region~\cite{Ablikim:2019hff, EcmsMea, EventFilter}. The cylindrical core of the BESIII detector covers 93\% of the full solid angle and consists of a helium-based multilayer drift chamber~(MDC), a time-of-flight system~(TOF), and a CsI(Tl) electromagnetic calorimeter~(EMC),
which are all enclosed in a superconducting solenoidal magnet providing a 1.0~T magnetic field.
The solenoid is supported by an octagonal flux-return yoke with resistive plate chamber muon identification modules interleaved with steel. 
The charged-particle momentum resolution at $1~{\rm GeV}/c$ is $0.5\%$, and the ${\rm d}E/{\rm d}x$
resolution is $6\%$ for electrons from Bhabha scattering. The EMC measures photon energies with a resolution of $2.5\%$ ($5\%$) at $1$~GeV in the barrel (end cap) region. The time resolution in the plastic scintillator TOF barrel region is 68~ps, while that in the end cap region was 110~ps. The end cap TOF system was upgraded in 2015 using multigap resistive plate chamber technology, providing a time resolution of
60~ps, which benefits about 83\% of the data used in this analysis~\cite{etof1,etof2,etof3}.

Monte Carlo (MC) simulated data samples produced with a {\sc geant4}-based~\cite{geant4} software package, which
includes the geometric description of the BESIII detector and the detector response, are used to determine detection efficiencies and to estimate backgrounds. The simulation models the beam energy spread and initial state radiation (ISR) in the $e^+e^-$ annihilations with the generator {\sc kkmc}~\cite{kkmc1,kkmc2}. The inclusive MC sample includes the production of the
$\psi(3686)$ resonance, the ISR production of the $J/\psi$, and the continuum processes incorporated in {\sc kkmc}~\cite{kkmc1,kkmc2}.

All particle decays are modeled with {\sc evtgen}~\cite{ref:evtgen1, ref:evtgen2} using branching fractions either taken from the Particle Data Group (PDG)~\cite{pdg}, when available, or otherwise estimated with {\sc lundcharm}~\cite{ref:lundcharm1, ref:lundcharm2}.
Final state radiation from charged final state particles is incorporated using the {\sc photos} package~\cite{photos2}.

Dedicated MC samples for the processes $\psi(3686)\to\gamma\eta(1405),\eta(1405)\to f_0(980)\pi^0\to\pi^+\pi^-\pi^0$, $\psi(3686)\to\gamma\eta_c, \eta_c \to \pi^+\pi^-\pi^0$ and $\psi(3686)\to\gamma\eta_c, \eta_c\to f_0(980)\pi^0\to\pi^+\pi^-\pi^0$ are generated to optimize event selection criteria and to determine the detection efficiencies.

\section{Event selection and data analysis}
\label{sec:selection}

For the selection of the process $\psi(3686)\to\gamma\pi^+\pi^-\pi^0$ with $\pi^0\to\gamma\gamma$, events with two oppositely charged  pion tracks and at least three photons are required.
Charged tracks detected in the MDC are required to be within $|\rm{cos\theta}|<0.93$, where $\theta$ is the polar angle relative to the MDC's symmetry axis ($z$-axis). Each track must also originate within $|V_{z}| < 10$\,cm and $|V_{xy}| < 1$\,cm of the interaction point (IP)

Particle identification (PID) combines measurements of d$E$/d$x$ and TOF information to form likelihoods $\mathcal{L}(h)$ for hypotheses $h=e, \pi$.  $\mathcal{L}(\pi) > \mathcal{L}(e)$ is required to  suppress the misidentification of electrons as pions.

Photon candidates are identified using isolated EMC showers with energy deposits exceeding 25~MeV (barrel, $|\cos \theta|< 0.80$) or 50~MeV (end cap, $0.86 <|\cos \theta|< 0.92$). 
To exclude track-induced backgrounds, showers must be separated from any charged track (measured from the IP) by at least 10 degrees. 
The EMC time must be within [0, 700]\,ns of the event start time to suppress electronic noise and showers unrelated to the event.

A four-constraint (4C) kinematic fit imposing energy-momentum conservation is applied to each $\pi^+\pi^-\gamma\gamma\gamma$ combination.
If more than three photon candidates are present in an event, the combination with the smallest $\chi^2_{\rm 4C}(3\gamma)$ is retained, and the $\chi^2_{\rm 4C}(3\gamma)$ is required to be less than 20. In order to reduce backgrounds from processes with two or four photons, additional 4C kinematic fits with 2- and 4-photon hypotheses must yield $\chi^2$ values greater than $\chi^2_{\rm 4C}(3\gamma)$.

Among the three selected photons, the $\gamma\gamma$ pair with the invariant mass closest to $m_{\pi^0}$ and satisfying $|M(\gamma\gamma)-m_{\piz}|<0.025$~GeV/$c^2$ is taken as the $\pi^0$ candidate,
where $m_{\pi^0}$ is the nominal mass of the $\piz$ from the PDG~\cite{pdg}. 
The remaining photon is assigned to be the radiative one. 
To remove backgrounds stemming from $\eta\to \gamma\gamma$ decays, the invariant mass of any $\gamma\gamma$ pair is required to be outside the $\eta$ region (0.52, 0.57)~GeV/$c^2$.

After applying the above mentioned  selection criteria, there are still some residual background events  from the processes  $\psi(3686)\to\pi^0 J/\psi (J/\psi\to\pi^+\pi^-\gamma$), $\psi(3686)\to\gamma \chi_{cJ} (  \chi_{cJ}\to \gamma J/\psi, J/\psi\to\pi^+\pi^-\gamma$) and $\psi(3686)\to\pi^+\pi^-\omega(\gamma\pi^0)$, which are removed by requiring $|M(\pi^+\pi^-\gamma)-m_{J/\psi}|>0.05$~GeV/$c^{2}$ and $|M(\pi^0\gamma)-m_{\omega}|>0.04$~GeV/$c^2$, respectively. The nominal masses of the $J/\psi$ and $\omega$, $m_{J/\psi}$ and $m_{\omega}$, are taken from  the PDG~\cite{pdg}.
In addition, an inclusive MC sample of 2.7 billion  $\psi(3686)$ events is used to investigate other potential backgrounds, which will be discussed in detail later.

\subsection{Analysis of $\eta(1405)\to f_0(980)\pi^0\to\pi^+\pi^-\pi^0$}

\begin{figure}[h]
    \centering
    \subfigure{
    \begin{overpic}[scale=0.43]{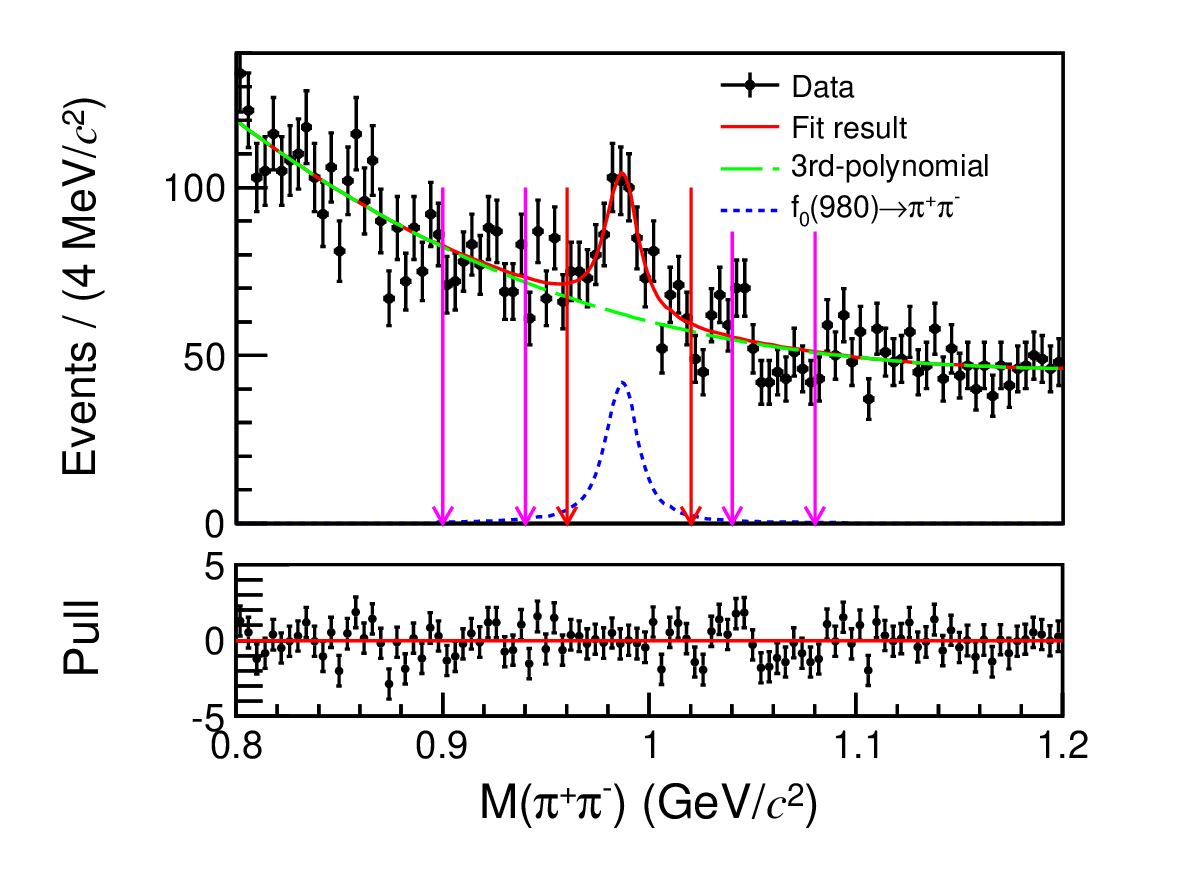}
    \put(80,65){\textbf{(a)}}
    \label{fig:f0980_fitresult}
    \end{overpic}
    }
    \\
    \vspace{-5mm}
    \subfigure{
    \begin{overpic}[scale=0.43]{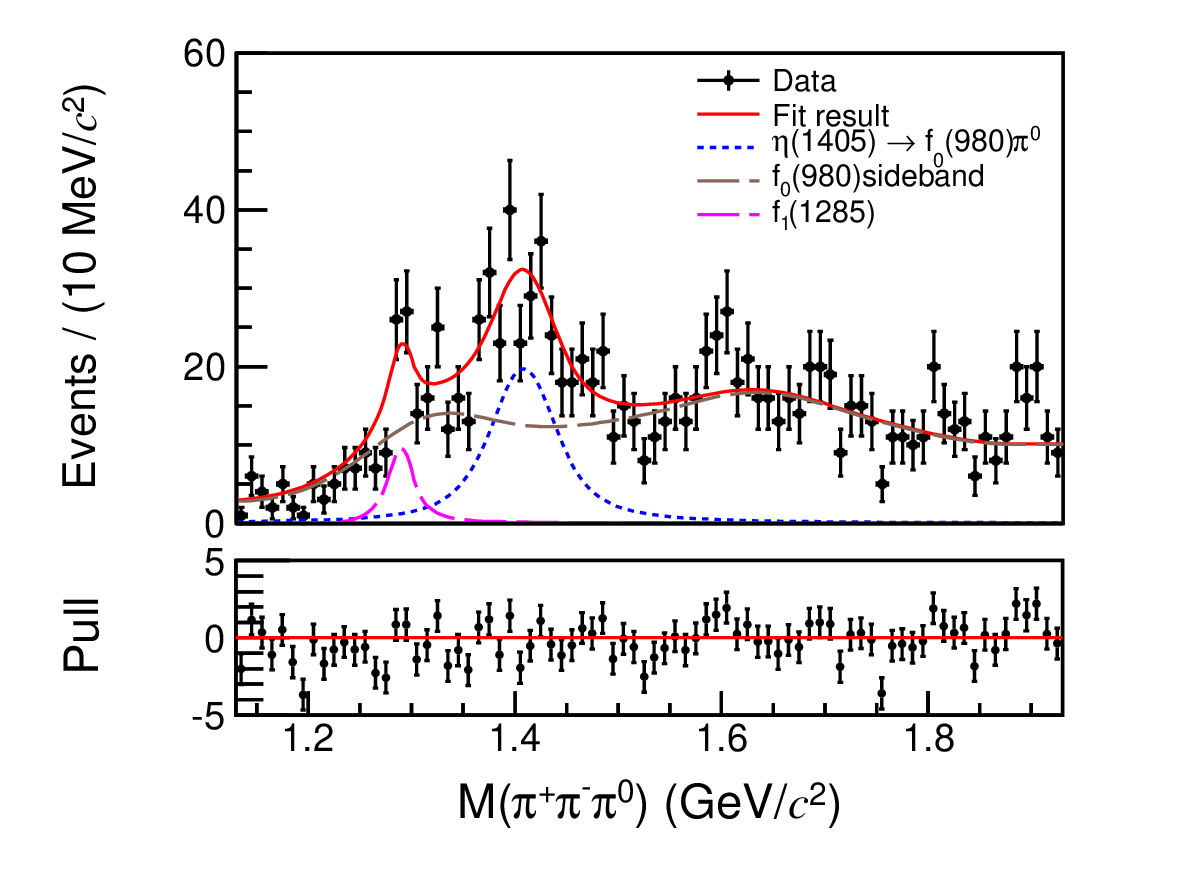}
    \put(80,65){\textbf{(b)}}
     \label{fig:eta1405_fitresult}
    \end{overpic}
    }
    \\
    \vspace{-3mm}
    \caption{(a) The $M(\pi^+\pi^-)$ distribution. The red arrows and pink arrows show the $f_0(980)$ signal region and the sideband regions, respectively. (b) The $M(\pi^+\pi^-\pi^0)$ distributions after selecting events in the $f_0(980)$ mass window. The black dots with error bars are data, the red solid line is the total fit result, the blue dashed line is the $\eta(1405)$ signal, the brown long-dashed line is the  $f_{0}(980)$ sideband, and the pink long-dashed line is the  $f_{1}(1285)$ peak.}
\end{figure}

After requiring the invariant mass of the $\pi^+\pi^-\pi^0$ system~($M(\pi^+\pi^-\pi^0)$) to be in the $\eta(1405)$ mass region ($(1.13, 1.93)$~GeV/$c^{2}$), the $\pi^+\pi^-$ invariant mass~($M(\pi^+\pi^-)$) is shown in Fig.~\ref{fig:f0980_fitresult}, where a clear $f_0(980)$ signal can be  observed. 
An unbinned maximum likelihood fit is performed to the $M(\pi^+\pi^-)$ distribution. In the fit, the $f_0(980)$ signal is  modeled with a Breit-Wigner function, with the mass and width  determined to be $\rm M=986.07\pm1.66$~MeV/$c^2$ and $\Gamma=21.23\pm5.68$~MeV/$c^2$, respectively. 
Furthermore, the smooth backgrounds, mainly from $\psi(3686)\to\gamma\chi_{cJ(J=0,1,2)}$ with $\chi_{cJ}\to\pi^+\pi^-\pi^0$ and $\psi(3686)\to\pi^+\pi^-\pi^0\pi^0$, are described using a third-order Chebychev polynomial.
A further requirement of $0.96$~GeV/$c^{2}$ $<M(\pi^+\pi^-)<1.02$~GeV/$c^{2}$ is used to select $f_0(980)$ candidates.

After applying the above selection criteria, the $M(\pi^+\pi^-\pi^0)$ distribution in the region of interest $(1.13, 1.93)$~GeV/$c^{2}$ is shown in Fig.~\ref{fig:eta1405_fitresult}, exhibiting a distinct $\eta(1405)$ signal.
To determine the $\eta(1405)$ signal yield, an unbinned maximum likelihood fit is performed to the $M(\pi^+\pi^-\pi^0)$ distribution. 
In the fit, the $\eta(1405)$ and $f_{1}(1285)$ signals are modeled using the MC-simulated shape convolved with a Gaussian function with floating parameters. 
The background is described by the $f_{0}(980)$ sidebands, 0.90~GeV/$c^{2}$ $<M(\pi^+\pi^-)<0.94$~GeV/$c^{2}$ and 1.04~GeV/$c^{2}$ $<M(\pi^+\pi^-)<1.08$~GeV/$c^{2}$, and the yield is fixed to the value estimated from the $M(\pi^+\pi^-)$ fit result. 
Assuming no interference, the fit result is shown in Fig.~\ref{fig:eta1405_fitresult}, and the $\eta(1405)$ signal yield is determined to be $N_{\rm signal}=195.33\pm22.43$, with a statistical significance of $10.9\sigma$, which is calculated from the likelihood difference ($\Delta$lnL=64.77) and the change in degrees of freedom ($\Delta$ndf=3) between fits with and without the $\eta(1405)$ signal component. 
The branching fraction is calculated as
\begin{equation}
    \centering
    \begin{aligned}
  &  \mathcal{B}(\psi(3686)\to\gamma\eta(1405)\to\gamma f_0(980)\pi^0\to\gamma\pi^+\pi^-\pi^0) \\&=\frac{N_{\rm signal}}{N_{\psi(3686)}\cdot\mathcal{B}(\pi^0\to\gamma\gamma)\cdot\epsilon}=(3.77\pm0.43)\times10^{-7},
\end{aligned}
\label{formula2}
\end{equation}
where the uncertainty is only statistical.
Here $N_{\psi(3686)}=(2712.4\pm14.3)\times10^6$ is the total number of $\psi(3686)$ events~\cite{ref::psip_event}, $\mathcal{B}(\pi^0\to\gamma\gamma)=(98.823\pm0.034)\%$ is the  branching fraction of $\pi^0\to \gamma\gamma$  from the PDG~\cite{pdg}, and $\epsilon=19.33\%$ is the detection efficiency estimated from dedicated MC simulation.
The yield of $f_{1}(1285)$ is $37.7\pm11.5$ and the selection efficiency is 19.11\% estimated with a phase space distributed MC simulation. The branching fraction is measured to be $\mathcal{B}(\psi(3686)\to\gamma f_1(1285)\to\gamma f_0(980)\pi^0\to\gamma\pi^+\pi^-\pi^0) = (7.36\pm2.25)\times 10^{-8}$, with statistical uncertainties only.  The statistical significance is only 2.9$\sigma$.

\subsection{Search for $\psi(3686)\to\gamma\eta_c,\eta_c\to\pi^+\pi^-\pi^0$}

\begin{figure}[]
    \centering
    \subfigure{
    \begin{overpic}[scale=0.43]{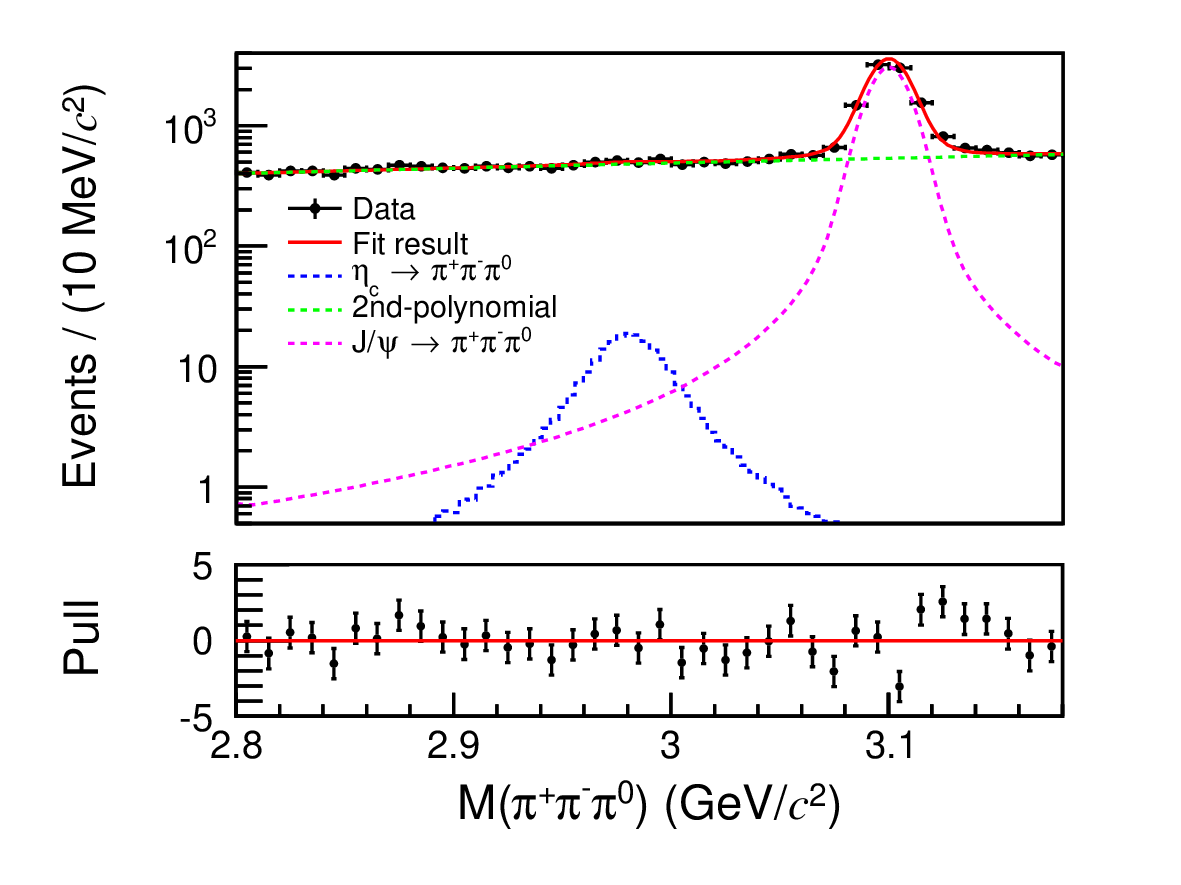}
    \put(80,65){\textbf{(a)}}
    \label{fig:etac_fit_log}
    \end{overpic}
    }
        \\
    \vspace{-5mm}
    \subfigure{
    \begin{overpic}[scale=0.43]{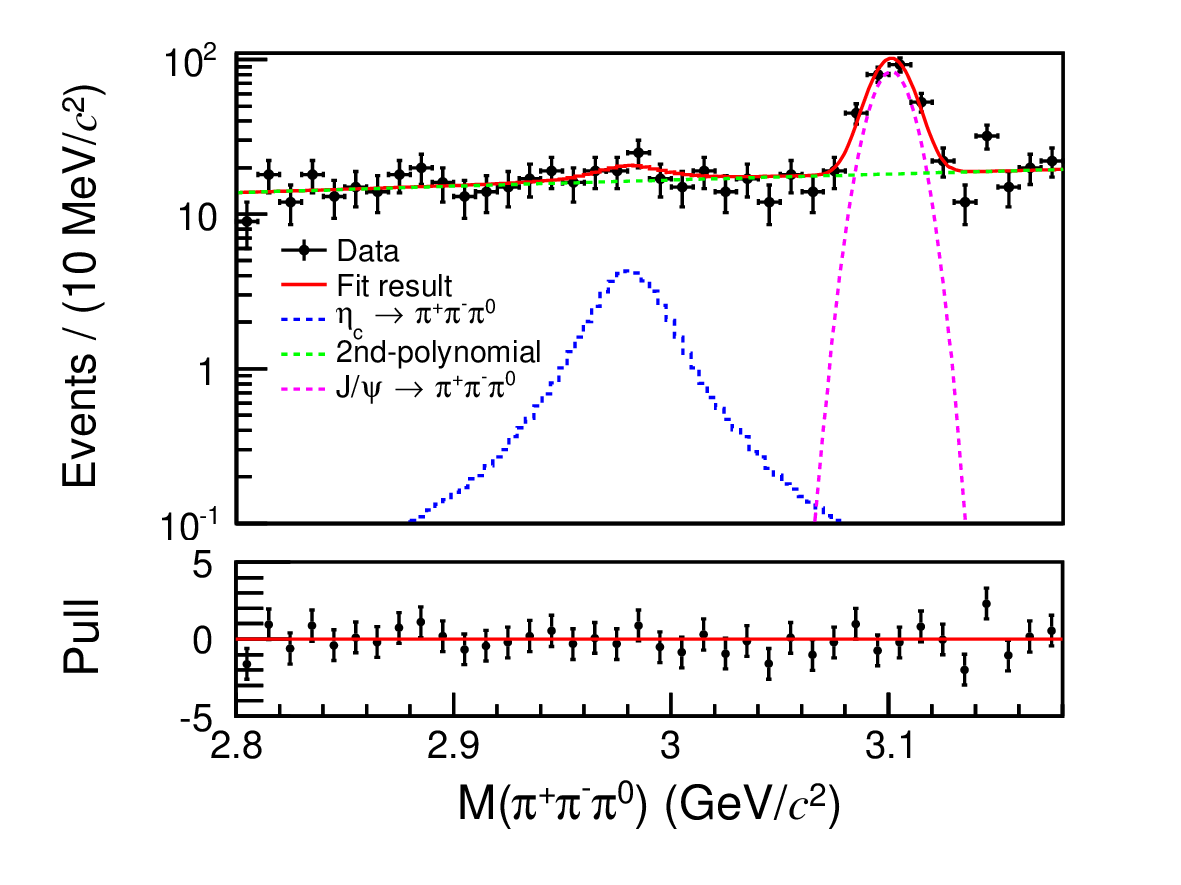}
    \put(80,65){\textbf{(b)}}
    \label{fig:etacf0_fitresult_log}
    \end{overpic}
    }
        \\
    \vspace{-5mm}
        \caption{Fits to the $M(\pi^+\pi^-\pi^0)$ distributions in the $\eta_c$ mass region (a) without and (b) with the requirement that $M(\pi^+\pi^-)$ is within the $f_0(980)$ mass window. 
        The black dots with error bars are  data, the red solid line is the total fits result, the blue dashed line is the $\eta_c$ signal shape, the pink dashed line is the $J/\psi$ background, and the green dashed line is  the combinatorial background.}
        \label{fig:etac_fitresult}
\end{figure}

To suppress additional background from $\psi(3686)\to\omega\pi^0$ with $\omega\to\pi^+\pi^-\pi^0$ in the $\eta_c$ mass region, a requirement of $0.7$~GeV/$c^{2}$ $<M(\pi^+\pi^-\gamma)<0.8$~GeV/$c^{2}$ is applied. The $M(\pi^+\pi^-\pi^0)$ distribution in the mass region (2.8, 3.18)~GeV/$c^{2}$ is shown in Fig.~\ref{fig:etac_fit_log}. There is no $\eta_c$ signal but a clear $J/\psi$ peak is observed. According to a topology analysis of the inclusive MC sample, these $J/\psi$ backgrounds come from 
$\psi(3686)\to\pi^0 J/\psi$, $\gamma\chi_{cJ(J=0,1,2)}$ with $\chi_{cJ}\to\gamma J/\psi$ and the QED process $e^+e^-\to\gamma_{\rm ISR} J/\psi$. The other combinatorial background events mainly come from $\psi(3686)\to\gamma\chi_{cJ}(\chi_{cJ}\to\pi^+\pi^-\pi^0)$ and $\psi(3686)\to \pi^+\pi^-\pi^0\pi^0$.  None of the backgrounds create a peak in the $\eta_c$ mass region.

The result of an unbinned maximum likelihood fit to the $M(\pi^+\pi^-\pi^0)$ distribution is shown in Fig.~\ref{fig:etac_fit_log}. 
In the fit, the potential $\eta_c$ signal is modeled using the MC-simulated shape, while the $J/\psi$ peak is described by a Breit-Wigner function convolved with a Gaussian resolution function and the combinatorial background is described by a second-order polynomial. 
The signal yield from this fit is  $95\pm87$ with a statistical significance of $1.3\sigma$.

Using the Bayesian method \cite{ref::bayesian}, the UL on the $\eta_c$ signal yield at the 90\% CL, defined as $N^{\rm UL}$, is determined. A scan of unbinned maximum likelihood fits are performed to the $M(\pi^+\pi^-\pi^0)$ distributions, testing the $\eta_c$ signal yield hypotheses from 0 to 200 in steps of 1. 
The probability density function (PDF) is obtained according to the normalized likelihood values of the fits, defined as $\mathcal{L}(N) = \exp(-[\mathcal{S}(N) - \mathcal{S}_{\min}])$, where $\mathcal{S}_{\rm min}$ is the minimum negative log-likelihood value obtained from the fits.
The 90\% CL upper limit on the signal yield, denoted as $N^{\rm UL}$, is determined by integrating the PDF up to the value that contains 90\% of the total probability,
\begin{equation}
    \frac{\int_{0}^{N^{\mathrm{UL}}} \mathcal{L}(N) dN}{\int_{0}^{\infty} \mathcal{L}(N) dN} = 0.9, 
\end{equation}
where $N$ is the expected number of signal events.
To account for the additive systematic uncertainties, alternative fits are performed by enlarging, reducing, and shifting the fit range. In these fits, second- or third-order polynomials are tested in describing the background. 
The $\eta_c$ signal shape is modeled using the MC-simulated shape convolved with a Gaussian function. 
In addition, individual MC samples are produced with different $\eta_c$ widths to describe the signal shape in alternative fits.
The maximum UL on the signal yield at the 90\% CL, $N^{\rm UL}$=221.7, is taken as the nominal value.

We also determine the UL of the signal yield of the subprocess via the $f_0(980)$ resonance.
To select the $f_0(980)$ signal region, an additional requirement of $0.96$~GeV/$c^{2}$ $<M(\pi^+\pi^-)<1.02$~GeV/$c^{2}$ is applied.
An unbinned maximum likelihood fit to the $M(\pi^+\pi^-\pi^0)$ distribution is performed, as presented in Fig.~\ref{fig:etacf0_fitresult_log}, where no significant $\eta_c$ signal is observed. 
The statistical significance of the $\eta_c$ signal is estimated to be $1.5\sigma$. 
Similarly, the $\eta_c$ signal is modeled using the MC-simulated shape, the $J/\psi$ background is described by a Breit-Wigner function convolved with a Gaussian resolution function, and the combinatorial background is described by a third-order polynomial. 
To obtain a conservative UL on the signal yield, alternative fits are performed by varying the $\eta_c$ width, signal shape, fit range and background shape. The maximum UL of the signal yield at the 90\% CL is found to be $N^{\rm UL}=59$.

\section{SYSTEMATIC UNCERTAINTY}
\label{sec:systematics}

Systematic uncertainties arising from various sources are investigated and summarized below. Assuming all the sources are independent, the total systematic uncertainty of each decay mode is calculated by adding the individual contributions in quadrature. 
Table \ref{tab:Sys} summarizes the systematic uncertainty results.

The uncertainties of the MDC tracking for each charged pion and the photon selection are investigated using a control sample of $J/\psi\to\rho\pi$~\cite{sys_track, sys_gamma}. The difference between data and MC simulation is 1.0\% per charged pion and also 1.0\% per photon. The uncertainty related to the pion PID is studied with the same sample~\cite{sys_pion}, and a 1.0\% difference per charged pion is assigned. Therefore, charged pions tracking, PID and photon selection contribute 2.0\%, 2.0\%, and 3.0\% uncertainties, respectively.

The uncertainties associated with the  kinematic fits come from the inconsistency of the track helix parameters between data and MC simulation.
Corrections are applied to improve agreement, especially in $\chi^2_{\rm 4C}$ distributions.
Following the method described in Ref.\cite{ref:kinematic_fit}, the differences in the selection efficiencies with and without the correction are taken as the systematic uncertainties, estimated to be 1.0\% for $\eta(1405)\to f_0(980)\pi^0\to\pi^+\pi^-\pi^0$, 2.3\% for $f_1(1285)\to f_0(980)\pi^0\to\pi^+\pi^-\pi^0$, 1.3\% for $\eta_c\to \pi^+\pi^-\pi^0$, and 1.2\% for $\eta_c\to f_0(980)\pi^0\to\pi^+\pi^-\pi^0$, respectively.
A 0.8\% uncertainty is assigned due to the $\pi^0$ mass window requirement, based on the study in Ref.~\cite{ref::gushanwork}.

In the fits to the $M(\pi^+\pi^-\pi^0)$ distributions for the $ \eta(1405) \to f_0(980)\pi^0 $ channel, the signal shape is represented by the MC-simulated shape convolved with a Gaussian function. To estimate the associated uncertainty, the fit is repeated using a Breit-Wigner function convolved with a Gaussian resolution function. The difference in the branching fractions between these two approaches,  2.9\% and 4.2\%, are taken as the associated uncertainties for the $ \eta(1405) \to f_0(980)\pi^0 $ channel and the $f_1(1285) \to f_0(980)\pi^0 $ channel, respectively.
    
The $f_0(980)$ width exhibits significant variation across decay channels. To estimate the related uncertainties, the values of the width in MC simulation are varied by one standard deviation, extracted from a fit to the  $M(\pi^+\pi^-)$ distribution in $\eta(1405)\to f_0(980)\pi^0\to\pi^+\pi^-\pi^0$ decay. The resulting shifts in the branching fraction, 4.0\% for $\eta(1405)\to f_0(980)\pi^0\to\pi^+\pi^-\pi^0$, 23.2\% for $f_1(1285)\to f_0(980)\pi^0\to\pi^+\pi^-\pi^0$, and 8.2\% for the UL of $\eta_c\to f_0(980)\pi^0\to\pi^+\pi^-\pi^0$, are taken as the systematic uncertainties.

The uncertainty of the background estimation comes from the choice of sideband ranges and the kernel bandwidth factor in the fit. 
Different $f_0(980)$ sideband ranges are used in the fit and the maximum differences in branching fraction, 2.7\% and 18.9\%, are taken as the systematic uncertainties for the $ \eta(1405) \to f_0(980)\pi^0 $ channel and the $f_1(1285) \to f_0(980)\pi^0 $ channel, respectively. 
The kernel bandwidth factors are expanded or reduced by 20\% in the fit and the maximum differences, 1.2\% and 9.3\%, are taken as the systematic uncertainties, for the $ \eta(1405) \to f_0(980)\pi^0 $ channel and the $f_1(1285) \to f_0(980)\pi^0 $ channel, respectively.

For $\eta(1405)\to f_0(980)\pi^0\to\pi^+\pi^-\pi^0$, the uncertainty associated with the fit range is checked by using different fit ranges. The maximum differences in the measured branching fraction, 0.5\% and 2.7\%, are taken as the systematic uncertainties for the $ \eta(1405) \to f_0(980)\pi^0 $ channel and the $f_1(1285) \to f_0(980)\pi^0 $ channel, respectively. 
The uncertainty associated with the signal shape mainly comes from the width of the $\eta(1405)$, which is estimated by varying the width by one standard deviation in the MC simulation according to the PDG value~\cite{pdg}. The maximum difference in the branching fraction of $\eta(1405)\to f_0(980)\pi^0 \to \pi^+\pi^-\pi^0$, 3.2\%, is taken as the systematic uncertainty.
For the ULs of  $\eta_c\to\pi^+\pi^-\pi^0$ and $\eta_c\to f_0(980)\pi^0\to\pi^+\pi^-\pi^0$, the additive uncertainties associated with the fit range, background shape, and signal shape are considered in the nominal ULs.

The uncertainty on the branching fraction $\pi^0\to\gamma\gamma$ $(98.823\pm0.034)\%$ from the PDG \cite{pdg} is 0.03\%.
The total number of $\psi(3686)$ events is determined to be $(2712.4\pm14.3)\times10^6$ in Ref.~\cite{ref::psip_event}, which contributes an uncertainty of 0.5\%.
\begin{table*}[]
   \centering
    \caption{Systematic uncertainties (in \%) on the measurements of  the branching fraction and ULs (multiplicative ones). Assuming the uncertainties are uncorrelated, the total uncertainty is the quadratic sum of the individual values. }
    \begin{tabular}{l|cccc}
   \hline
   \hline
  Source & $\eta(1405)$ &$f_1(1285)$ &$\eta_c$  & $\eta_c\to f_0(980)\pi^0$\\ 
\hline
  Tracking                     & 2.0 & 2.0   & 2.0  & 2.0\\ 
  Photon detection             & 3.0 & 3.0   & 3.0  & 3.0\\ 
  PID                          & 2.0 & 2.0   & 2.0  & 2.0\\ 
  Kinematic fit                & 1.0 & 2.3   & 1.3   & 1.2\\
  $\pi^0$ mass window          & 0.8 & 0.8   & 0.8  & 0.8\\ 
  Width of $f_0 (980)$         & 4.0 & 23.2     & -  & 8.2\\ 
   Signal shape   & 2.9 & 4.2     & -  & -  \\
  Fitting range                & 0.5 & 2.7     & -    & -\\ 
  Background estimation        & 3.0 & 21.1     & -    & -\\ 
  Width of $ \eta(1405)$       & 3.2 & -    & -    & -     \\ 
  $\mathcal{B}$($\pi^0\to\gamma\gamma$)   & 0.03 & 0.03  & 0.03 & 0.03 \\ 
  Number of $\psi$(3686) events      & 0.5  & 0.5  & 0.5  & 0.5\\ \hline
  Total                        & 7.9  & 32.1   & 4.4 & 9.3 \\
   \hline
   \hline
   \end{tabular}
    \label{tab:Sys}
\end{table*}

\section{Result}
Considering the systematic uncertainty, the branching fraction of the decay $\psi(3686)\to\gamma\eta(1405)\to\gamma f_0(980)\pi^0\to\gamma\pi^+\pi^-\pi^0$ is determined using Eq.~\eqref{formula2} to be
\begin{equation}
    \begin{aligned}
  &\mathcal{B}(\psi(3686)\to\gamma\eta(1405)\to\gamma f_0(980)\pi^0\to\gamma\pi^+\pi^-\pi^0) 
  \\&=(3.77\pm0.43\pm0.30)\times10^{-7}.
    \end{aligned}
\end{equation}
The branching fraction of the decay $\psi(3686)\to\gamma f_1(1285)\to\gamma f_0(980)\pi^0\to\gamma\pi^+\pi^-\pi^0$ is determined to be 
\begin{equation}
    \begin{aligned}
    &\mathcal{B}(\psi(3686)\to\gamma f_1(1285)\to\gamma f_0(980)\pi^0\to\gamma\pi^+\pi^-\pi^0) 
    \\ &= (7.36\pm2.25\pm2.36)\times 10^{-8}.
    \end{aligned}
\end{equation}
The $\eta(1405)\to f_0(980)\pi^0$ signal could arise from $\eta(1405)\to a_0(980)\pi^0$ via $a_0(980)-f_0(980)$ mixing. 
The $a_0(980)-f_0(980)$ mixing intensity has been measured by \besiii to be $(0.40\pm0.07\pm0.14)\times10^{-2}$ via $\chi_{c1}\to f_0(980)\pi^0\to\pi^+\pi^-\pi^0$ and $\chi_{c1}\to a_0(980)\pi^0\to\eta\pi^0\pi^0$ \cite{ref::mixing}. 
In combination with the ratio $\Gamma(\eta(1405)\to a_{0}(980)\pi)/\Gamma(\eta(1405)\to\eta\pi\pi)=0.56\pm0.04\pm0.03$ and the branching fraction $\mathcal{B}(\psi(3686)\to\gamma\eta(1405)\to\gamma\eta\pi^+\pi^-)=(0.36\pm0.25\pm0.05)\times10^{-4}$ taken from the PDG \cite{pdg}, the expected branching fraction $\mathcal{B}(\psi(3686)\to\gamma\eta(1405)\to\gamma a_{0}(980)\pi^0\to\gamma\pi^0\eta\pi^0)$ is calculated to be $(1.01\pm0.71)\times10^{-5}$. 
Therefore, the branching fraction of $\psi(3686)\to\gamma\eta(1405)\to\gamma a_0(980)\pi^0\to\gamma f_0(980)\pi^0\to\gamma\pi^+\pi^-\pi^0$ via $a_0(980)-f_0(980)$ mixing is expected to be $(4.04\pm3.24)\times10^{-8}$, which is significantly smaller than the result measured by this analysis. This indicates that $a_0(980)-f_0(980)$ mixing is insufficient to explain the branching fraction of $\eta(1405)\to f_0(980)\pi^0$.

\begin{figure}[]
    \centering
    \subfigure{
    \begin{overpic}[scale=0.4]{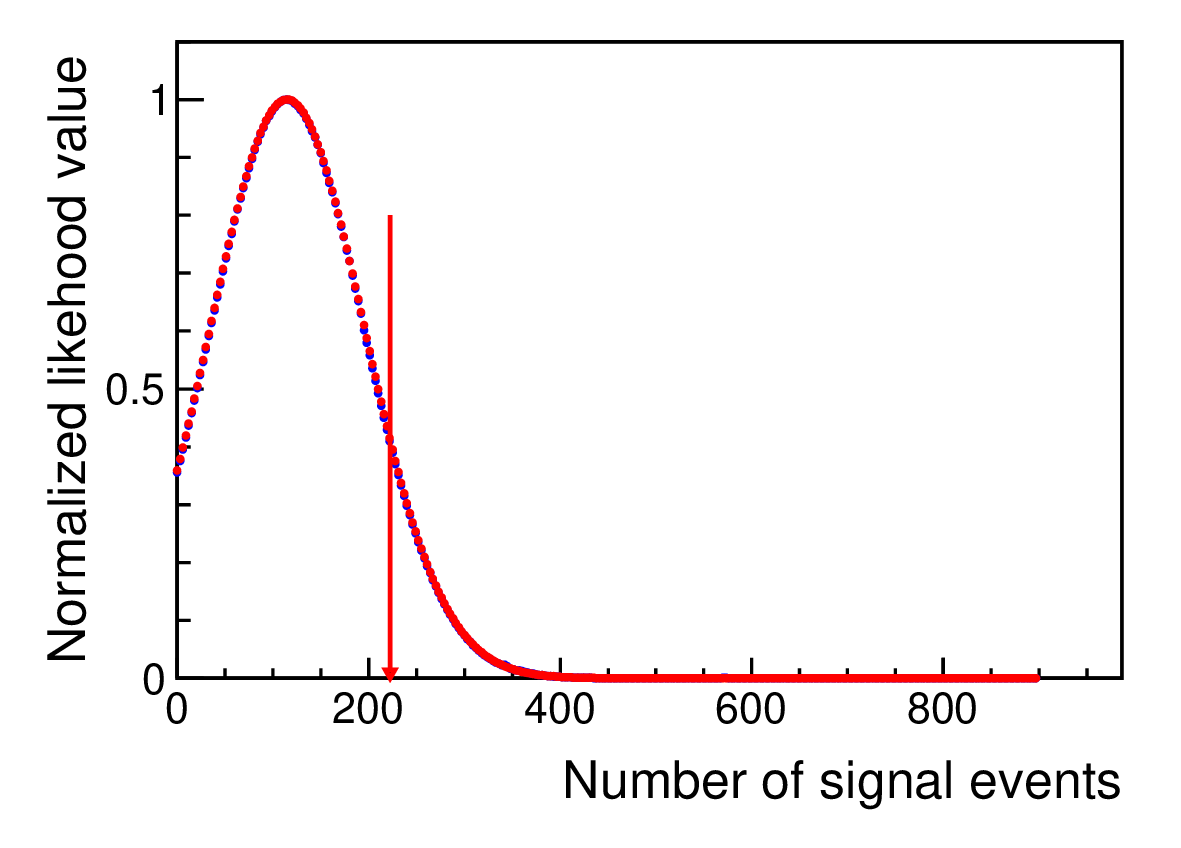}
    \put(80,60){\textbf{(a)}}
        \label{fig:etacf0_smearCL}
    \end{overpic}
    }
        \\
    \vspace{-5mm}
    \subfigure{
    \begin{overpic}[scale=0.4]{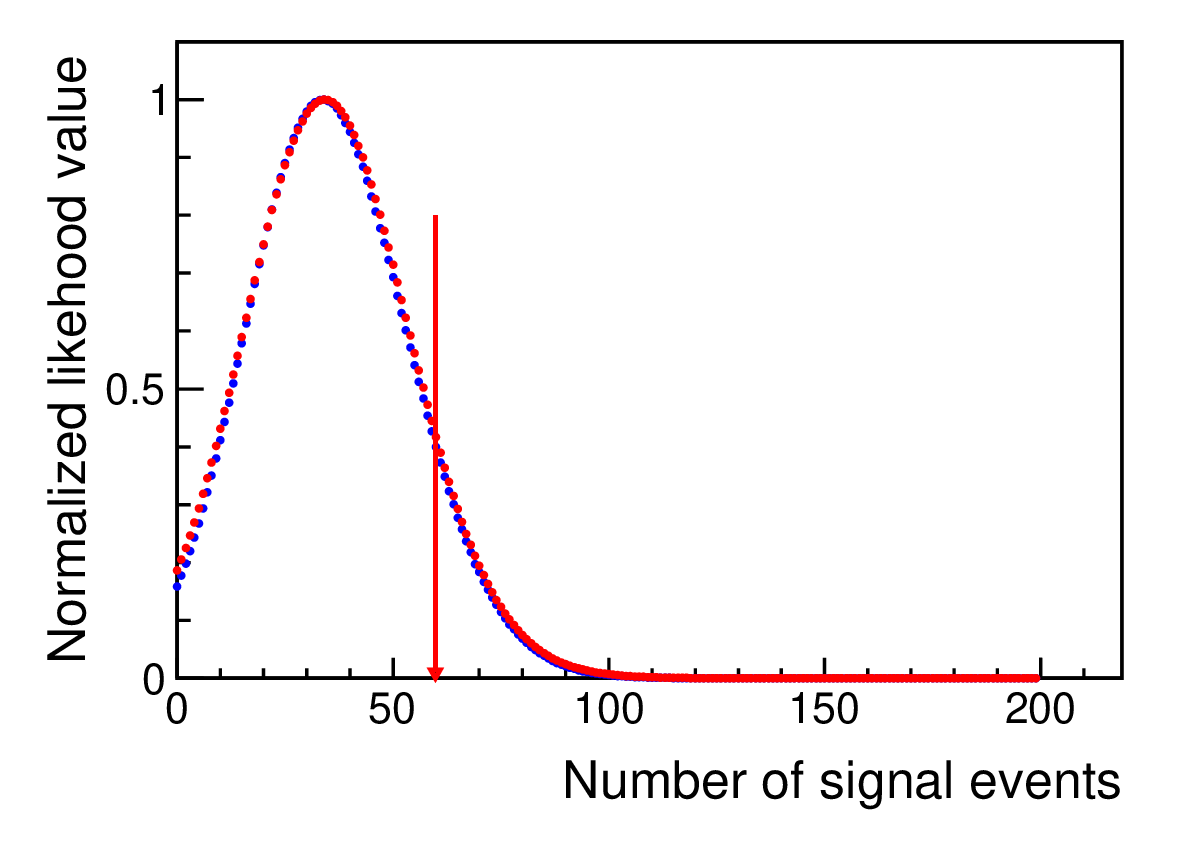}
    \put(80,60){\textbf{(b)}}
    \label{fig:etac_smearCL}
    \end{overpic}
    }
        \\
    \vspace{-5mm}
 \caption{Normalized likelihood distribution before (blue dots) and after (red dots) considering the multiplication uncertainties for (a) $\eta_c\to\pi^+\pi^-\pi^0$ and (b) $\eta_c\to f_0(980)\pi^0$.}
    \label{fig:smearCL}
\end{figure}

For the $\eta_c\to\pi^+\pi^-\pi^0$ and $\eta_c\to f_0(980)\pi^0$, 
the UL on the branching fraction is conservatively estimated by incorporating multiplicative systematic uncertainties into the PDF using a convolution method~\cite{smear}:
\begin{equation}
L^{\prime}(N)=\int_0^1 L\left(\frac{\mathcal{S}}{\hat{\mathcal{S}}} N\right) \exp \left[-\frac{(\mathcal{S}-\hat{\mathcal{S}})^2}{2 \sigma_S^2}\right] d \mathcal{S},
\end{equation}
where $S$ and $\hat{\mathcal{S}} $ are the detection efficiency and the nominal efficiency, respectively; $L$ represents the likelihood distribution without taking into account the systematic uncertainty; and $ \sigma_S $ denotes the multiplicative systematic uncertainty.
Figure~\ref{fig:smearCL} illustrates the normalized likelihood distributions before and after accounting for the multiplicative systematic uncertainty.

With the convolved PDF, the UL on the signal yield at 90\% CL are $N^{\rm UL}$=222.2 for  $\eta_c\to\pi^+\pi^-\pi^0$ 
and $N^{\rm UL}$=59.7 for  $\eta_c\to f_0(980)\pi^0$, respectively. The corresponding ULs on the branching fraction of $\psi(3686)\to\gamma\eta_c,\eta_c\to\pi^+\pi^-\pi^0$ and $\psi(3686)\to\gamma\eta_c,\eta_c\to f_0(980)\pi^0\to\pi^+\pi^-\pi^0$ are calculated to be
\begin{equation}
 \begin{aligned}
 &\mathcal{B}^{\rm UL}(\psi(3686)\to\gamma\eta_c,\eta_c\to\pi^+\pi^-\pi^0)=3.09\times10^{-7},
\end{aligned}
\end{equation}

\begin{equation}
 \begin{aligned}
  &\mathcal{B}^{\rm UL}(\psi(3686)\to\gamma\eta_c,\eta_c\to f_0(980)\pi^0\to\pi^+\pi^-\pi^0)
  \\& =7.97\times10^{-8}.
\end{aligned}
\end{equation}

\section{Summary}
Using a sample of $(2712.4\pm14.3)\times10^6$ $\psi(3686)$ events collected with the BESIII detector, the decay $\psi(3686)\to\gamma\eta(1405)\to\gamma f_0(980)\pi^0\to\gamma\pi^+\pi^-\pi^0$ is observed for the first time. The branching fraction is measured to be $(3.77\pm0.43\pm0.30)\times10^{-7}$, where the first uncertainty is statistical and the second is systematic. The branching fraction cannot be explained solely by $f_0(980)-a_0(980)$ mixing. Additional contributions, such as those from the triangle singularity mechanism, are necessary. 
By combining the branching fraction of $J/\psi\to\gamma\eta(1405)\to\gamma f_0(980)\pi^0\to\gamma\pi^+\pi^-\pi^0$, $(1.5\pm0.11\pm0.11)\times10^{-5}$~\cite{pdg}, we determine the ratio 
$
\frac{\mathcal{B}(\psi(3686)\to\gamma\eta(1405))}{\mathcal{B}(J/\psi\to\gamma\eta(1405))}=(2.51\pm 0.44)\%,
$
which indicates that this process violates the ``12\% rule''. 

Although the uncertainty is somewhat large, evidence for the process  $ f_1(1285)\to f_0(980)\pi^0$ in  $\psi(3686)$ radiative decays is reported for the first time. By combining the branching fraction of $J/\psi\to\gamma f_1(1285)\to\gamma f_0(980)\pi^0\to\gamma\pi^+\pi^-\pi^0$, $(9.99\pm3.00\pm1.03)\times10^{-7}$~\cite{pdg}, we determine the ratio 
$
\frac{\mathcal{B}(\psi(3686)\to\gamma f_1(1285))}{\mathcal{B}(J/\psi\to\gamma f_1(1285))}=(13.57\pm 7.41)\%,
$
which indicates that this process satisfies the ``12\% rule''.

In addition, we search for the decay $\psi(3686)\to\gamma\eta_c\to\gamma\pi^+\pi^-\pi^0(f_0(980)\to\pi^+\pi^-)$, but no clear $\eta_c$ signal is observed in either the $\pi^+\pi^-\pi^0$ or $f_0(980)\pi^0$ mass spectra. The ULs on the branching fractions at the 90\% CL are calculated to be $3.09\times10^{-7}$ for $\psi(3686)\to\gamma\eta_c,\eta_c\to\pi^+\pi^-\pi^0$ and $7.97\times10^{-8}$ for $\psi(3686)\to\gamma\eta_c,\eta_c\to f_0(980)\pi^0\to\pi^+\pi^-\pi^0)$, respectively. Compared with previous results in the PDG~\cite{pdg}, the obtained ULs are improved by a factor of 5.2.

\section*{ACKNOWLEDGMENTS}
The BESIII Collaboration thanks the staff of BEPCII~\cite{BEPCII} and the IHEP computing center for their strong support. This work is supported in part by National Key R\&D Program of China under Contracts No. 2025YFA1613900, No. 2023YFA1606000, No. 2023YFA1606704; National Natural Science Foundation of China (NSFC) under Contracts No. 11635010, No. 11935015, No. 11935016, No. 11935018, No. 12025502, No. 12035009, No. 12035013, No. 12061131003, No. 12192260, No. 12192261, No. 12192262, No. 12192263, No. 12192264, No. 12192265, No. 12221005, No. 12225509, No. 12235017, No. 12361141819; the Chinese Academy of Sciences (CAS) Large-Scale Scientific Facility Program; CAS under Contract No. YSBR-101; 100 Talents Program of CAS; The Institute of Nuclear and Particle Physics (INPAC) and Shanghai Key Laboratory for Particle Physics and Cosmology; ERC under Contract No. 758462; German Research Foundation DFG under Contract No. FOR5327; Istituto Nazionale di Fisica Nucleare, Italy; Knut and Alice Wallenberg Foundation under Contracts No. 2021.0174, No. 2021.0299; Ministry of Development of Turkey under Contract No. DPT2006K-120470; National Research Foundation of Korea under Contract No. NRF-2022R1A2C1092335; National Science and Technology fund of Mongolia; Polish National Science Centre under Contract No. 2024/53/B/ST2/00975; STFC (United Kingdom); Swedish Research Council under Contract No. 2019.04595; U. S. Department of Energy under Contract No. DE-FG02-05ER41374; This paper is also
supported by the Guangdong Basic and Applied Basic Research Foundation under Contract No. 2024A1515012416.

\section*{DATA AVAILABILITY}
The data that support the findings of this article are not publicly available. The data are available from the authors upon reasonable request.

\bibliographystyle{apsrev4-1}

\bibliography{pipi3gam.bib}

\end{document}